# New Quantum Spin Hall Insulator in Two-dimensional MoS$_2$ with Periodically Distributed Pores

Peng-Fei Liu,[†,‡] Liujiang Zhou,[*,⊥,&] Thomas Frauenheim,[⊥] Li-Ming Wu[*,†]

[†] *State Key Laboratory of Structural Chemistry, Fujian Institute of Research on the Structure of Matter, Chinese Academy of Sciences, Fuzhou, Fujian 350002, People's Republic of China*

[‡] *University of Chinese Academy of Sciences, Beijing 100039, People's Republic of China*

[⊥] *Bremen Center for Computational Materials Science, University of Bremen, Am Falturm 1, 28359 Bremen, Germany*

[&] *Max Planck Institute for Chemical Physics of Solids, Noethnitzer Strasse 40, 01187 Dresden, Germany*

*To whom correspondence should be addressed

E-mail: liujiang86@gmail.com, E-mail: liming_wu@fjirsm.ac.cn

## ABSTRACT

MoS$_2$, one of transition metal dichalcogenides (TMDs), has caused a lot of attentions for its excellent semiconductor characteristics and potential applications. Here, based on the density functional theory methods, we predict a novel two-dimension (2D) quantum spin hall (QSH) insulator in the porous allotrope of monolayer MoS$_2$ (g-MoS$_2$), consisting of MoS$_2$ square and hexagon. The g-MoS$_2$ has a nontrivial gap as large as 109 meV, comparable with previous reported 1T′-MoS$_2$ (80 meV), so-MoS$_2$ (25 meV). We demonstrate that the origin of 2D QSH effect in g-MoS$_2$ originates from the pure $d-d$ band interaction, different from conventional band inversion between $s-p$, $p-p$ or $d-p$ orbitals. Such new polymorph greatly enriches the TMDs family and its stabilities are confirmed by phonon spectrum analysis. In particular, porous structure also endows it potential application in efficient gas separation and energy storage.

# TOC Graphic

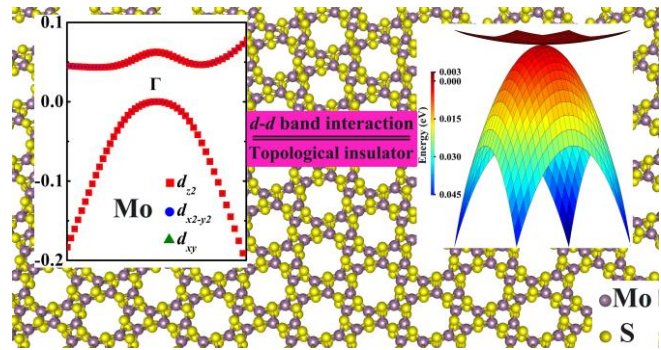

    Two-dimensional topological insulators (TIs), featured with gapless boundary states, which is protected by time-reversal symmetry with characteristic spin texture, completely refreshed our minds and brought a new revolution to material science due to their unparalleled electronic properties as well as promising applications in dissipationless electronic devices.[1-4] The edge states in two-dimensional (2D) TIs, characterized with quantum spin Hall (QSH) states, are more robust against nonmagnetic impurities than in 3D TIs and thus are better suited for coherent spin transport related applications. Impressive progress in searching for desired 2D TIs leads to successful findings of monolayers or few-layer van der Waals crystals, such as monolayer graphene,[3,4] silicene,[5,6] germanene,[7] stanene,[8] BiSb Alloys,[9] BiTeCl,[10] V$_2$−VI$_3$ family compounds (Bi$_2$Se$_3$, Bi$_2$Te$_3$ and Sb$_2$Te$_3$).[11] However, up to now the well-known QSH insulators, including HgTe/CdTe,[12] InAs/GaSb quantum wells,[13] were experimentally observed only at very low temperatures and ultrahigh vacuum due to weak spin−orbit coupling (SOC). To expand and advance practical application of two-dimensional (2D) TIs at room temperature, it is desired to design and search for new TIs to overcome the thermal disturbance. Intensive efforts have been devoted to engineer QSH insulators via first-principles method, and thus a crowd of candidates, including ZrTe$_5$,[14] ZrBr,[15] 2D III−Bi compounds,[16] methyl-munctionalized compounds (Bi Bilayer,[17] GeCH$_3$[18]), halide-munctionalized 2D materials(such as

X2-GeSn,[19] GeX,[20] decorated stanene,[21] Bi$_4$F$_4$,[22] Bi$_4$Br$_4$,[23] BiX/SbX monolayers,[24] chloridized gallium bismuthide[25] and so on), are predicted to be new TIs. Even though some of the predicted QSH insulators[17,22,25] show large band gap enough for room temperature (RT) applications, all of them still have not been obtained experimentally. Therefore, desirable materials preferably with large bulk gaps are still lacking and deserve to be explored in experiment and theory for realistic RT applications.

MoS$_2$, a typical example of the most studied transition metal dichalcogenides (TMDs), has attracted a lot of attention for its promising semiconductor characteristics and potential applications in in electronic, optical,[26] catalytic,[27] and lubricant properties.[28] Meanwhile, monolayer MoS$_2$ nanostructures exhibit even more intriguing properties, due to its intriguing properties by virtue of the quantum size effect, such as the strong photoluminescence,[29] moderation direct band gap (~1.8eV)[30,31] and relatively high mobility rate[30] and high on/off ratio.[33,34] Theoretical and experimental studies of monolayer MoS$_2$ have revealed its unlimited potentials for future applications in nanoelectronics.

Recently, the prediction of MoS$_2$ in the square-octagonal lattice (so-MoS$_2$)[35-37] was found to exhibit gapless band structure with Dirac fermions at Γ point. Distinct from the conventional $p-p$ band inversion in graphene, so-MoS$_2$ holds the pure $d-d$ band interaction and possess a Fermi velocity (2.3-2.4 × 10$^6$ m/s),[35] comparable to that of graphene.[2,3] Besides, lattice distortion can induce an intrinsic band inversion between chalcogenide-$p$ and metal-$d$ bands in 1T′-MoS$_2$,[38] a new class of large-gap QSH insulators in two-dimensional transition metal dichalcogenides. These researches here greatly enrich the physical properties of TMDs and provide new guidance for engineering high-performance TIs materials. Meanwhile, intensive efforts on the grain boundaries of monolayer MoS$_2$[39-41] have proved that four-, five-, seven-, eight-membered rings, as well as six-membered rings can coexist in monolayer MoS$_2$, showing the structure diversity. Motivated by above phenomenon, we design a new allotrope of MoS$_2$ monolayer composed of repeated special MoS$_2$ square and hexagon units (denoted as g-MoS$_2$). Based on density functional theory (DFT) method, we theoretically study the structure, stability, and electronic properties of the atom-thick

MoS$_2$ monolayer. This new kind of QSH insulator is verified by topological edge states and Z$_2$ topological invariant. The unique arrangements of six- and four-rings provide vital insight into physical properties of TMDs and open up a viable approach to design 2D topological materials for realistic applications.

Utilizing density functional theory (DFT) as implemented in the Vienna *ab-initio* simulation package (VASP),[42] we investigate the equilibrium structure, stability, and electronic properties of the predicted structure. All-electron projector augmented wave method[43] was used for the ionic cores and the generalized gradient approximation for the exchange-correlation potential.[44] The reciprocal space was sampled with 0.03 Å$^{-1}$ spacing in the Monkhorst-Pack scheme for structure optimization, and denser k-point grids with 0.01 Å$^{-1}$ spacing were adopted for electronic properties calculation. We used a mesh cutoff energy of 500 eV to determine the self-consistent charge density. All geometry structures were relaxed until the Hellmann-Feynman force on atoms is less than 0.01 eV/Å and the total energy variation is less than $1.0 \times 10^{-6}$ eV. The screened exchange hybrid density functional by Heyd-Scuseria-Ernzerhof (HSE06)[45,46] was adopted to further correct the electronic structure. A vacuum space of 15 Å along the z direction was used to avoid interactions between adjacent layers. The phonon calculations were carried out by using the density functional perturbation theory (DFPT)[47] as implemented in the PHONOPY code[48] combined with the VASP. To verify the stability of the system at elevated temperatures, the *ab-initio* molecular dynamic (MD) simulations were performed using the Nosé algorithm[49] in the NVT ensemble at 500K and 1000K respectively. The VESTA software[50] was used for visualization and plot.

The novel plane MoS$_2$ akin to graphenylene (Fig. S1c), is composed of repeated special MoS$_2$ square and hexagon units (denoted as g-MoS$_2$).[51] The optimized g-MoS$_2$ contains two six-membered MoS$_2$ rings connected by a four-membered MoS$_2$ unit, crystallizing in the hexagonal space group, $P\bar{6}m$ (no. 175), with $a = b = 8.803$ Å (Fig. 1a). Such a g-MoS$_2$ holds three distinct Mo-S bond lengths varying from 2.39 Å to 2.46 Å to 2.47 Å (Fig. 1c), revealing a severely structural distortion originated from the square-hexagon topology. Similar to other monolayer MoS$_2$, the g-MoS$_2$ layer with the covalently bonded S-Mo-S atoms has a Mo atomic plane layer sandwiched between two S atomic layers. All Mo atoms are hexa-coordinated octahedrally with six nearest neighbor S atoms. As illustrated in Fig. 1, we have calculated the thickness of the monolayer g-MoS$_2$ by simply measuring distance between the top and bottom S atomic layers. Compared to previous work,[35] the thickness of three MoS$_2$ allotropes, h-MoS$_2$ so-MoS$_2$ and g-MoS$_2$, decrease from 3.13 to 3.12 to 3.11 Å, respectively.

To evaluate the stability of this structure, we first computed the formation energy with respect to isolated atoms, defined as below:

$$E_{form} = \frac{E_{total} - n_{Mo}E_{Mo} - n_S E_S}{n_{Mo} + n_S} \qquad (1)$$

Where $E_{total}$ is the total energy of monolayer MoS$_2$, $E_{Mo}$ and $E_S$ were the energies an isolated Mo and S atom, respectively. Our result shows that g-MoS$_2$ has a formation energy of −4.79 eV, indicating it being energetically favorable. As a comparison, we also calculated the formation energy of h-MoS$_2$ and so-MoS$_2$, which is −5.08 eV and −4.80 eV, same to the reported value,[35] indicating g-MoS$_2$ and so-MoS$_2$ hold almost the same structural stability. Furthermore, the phonon band structure and vibrational density of states, shown in Fig. S2a, were calculated to study the dynamic stability and structural rigidity of the g-MoS$_2$. At first glance of the phonon dispersion, there are the linearly crossing phonon branches at the K symmetry point, which often appears in the hexagonal crystal. No imaginary phonon frequencies are observed in the Brillouin zone, indicating inherent dynamic stability of a crystal at low temperature. The presence of quite high eigenvalues in optical phonon is another direct indication for the structural stability. We find the highest frequency in phonon spectrum of g-MoS$_2$ reaches up to 442 cm$^{-1}$, close to that of so-MoS$_2$ (444 cm$^{-1}$), slightly lower than that of h-MoS$_2$ (462 cm$^{-1}$), reflecting a similar stability. The fluctuations of total energy with simulation time are plotted in Fig. S2b. After 5000 steps at 500 K, we found that there is no obvious structure destruction, and that the average value of total energy remains nearly constant during the whole simulation scale, confirming g-MoS$_2$ being at least thermally stable at room temperature.

The calculated band structure, presented in Fig. 1d, shows zero band gap with the valence band maximum (VBM) and conduction band minimum (CBM) touching at the Γ point. The shapes of the conduction and valence bands (VB and CB) are presented in Fig. 1e. The CB is flattened, while VB shows a considerable radian, indicating the coexistence of heavy electrons and light holes. The effective mass can be estimated by using the following equation at the bands extrema:

$$\frac{1}{m^*} = \frac{1}{\hbar^2} \frac{\partial^2 E_n(k)}{\partial k^2} \tag{2}$$

Our calculations show that the carrier effective masses are 8.54 $m_e$ (electron) and 0.39 $m_e$ (hole) along Γ − K direction at Γ point, respectively. Such a dramatic difference in effective mass is useful for selectively injecting or emitting holes or electrons, rendering the materials huge application prospects in nanoelectronic devices.

It is amazing that so-$MoS_2$ and g-$MoS_2$ are both gapless and have heavy carriers, but so-$MoS_2$ possesses unique massless Dirac structure. To explain this peculiar phenomenon, total DOS and projections DOS for g-$MoS_2$ so-$MoS_2$ and h-$MoS_2$ are calculated and plotted in Fig. 2. In h-$MoS_2$, the frontier orbitals mainly originate from the Mo−d orbitals hybridized with visible contributions from the S-p orbitals. In the DOS of g-$MoS_2$ and so-$MoS_2$, we can see the orbitals around Fermi level mainly come from Mo−d orbitals with negligible components from S−p orbitals, holding the zero DOS and thus showing zero-gap at Fermi level. In detail, in the case of h-$MoS_2$, the VB can be primarily ascribed to the $d_{x2-y2}$ and $d_{z2}$ orbitals while CB to $d_{z2}$ orbitals. As for so-$MoS_2$ and g-$MoS_2$, the basic units are totally reconstructed, and the electron orbital contributions around Fermi level have also varied. In the case of so-$MoS_2$, the VB and CB+1 are mainly composed of the Mo−$d_{x2-y2}$ orbitals, possessing a linear energy–momentum dispersion characterized by the Dirac structure, while the CB and VB−1 mainly composed of the Mo−$d_{z2}$ orbitals are almost flat. Meanwhile in the case of g-$MoS_2$, the CB and VB are primarily constituted of the Mo−$d_{z2}$ orbitals, having parabolic energy–momentum dispersion relations. So, the contributions from S−p electrons are almost negligible for g-$MoS_2$ and so-$MoS_2$, but indispensable for h-$MoS_2$.

To further quantify the contributions to CB and VB for g-$MoS_2$, we have calculated the three main d orbitals contributions to VB and CB along K-M-Γ-K in the Brillouin zone. In Fig. 3a, we can see the VB and CB are mostly constituted of $d_{z2}$ orbital with small contributions from Mo−$d_{xy}$ and −$d_{x2-y2}$ orbitals. Specifically in VB region, the $d_{xy}$, $d_{z2}$ and $d_{x2-y2}$ orbitals show no big fluctuations and the subordinate components of VB are $d_{x2-y2}$. While in CB region, the $d_{xy}$, $d_{z2}$ and $d_{x2-y2}$ orbitals undulate severely in the Brillouin zone and $d_{xy}$ orbital becomes the second part of CB around Γ point. To give an ocular explanation on the nature of the zero band gap of g-MoS$_2$, VBM charge density contours at Γ point are calculated and shown in Fig. 3b. We can clearly see that the frontier orbitals at Fermi level mainly originate from the Mo-$d_{z2}$ orbital hybridized with small amounts of $d_{xy}$ and $d_{x2-y2}$ orbitals, further verifying above analysis.

As is well-known, the four pear-shaped lobes of in-plane $d_{x2-y2}$ orbitals, shown in Fig. S4, spread in the x−y plane in real place, fully compatible with the square lattice symmetry. As a result, a long-range coherence is realized between the $d_{x2-y2}$ orbital and the square crystal lattice, leading to Dirac cone feature in so-MoS$_2$.[35] On the contrary, the two pear-shaped regions of out-of-plane $d_{z2}$ orbitals are localized and placed symmetrically along the c axis, bringing about the heavy fermions to h-MoS$_2$, so-MoS$_2$ and g-MoS$_2$. Overall, distinctive electronic properties in the three two-dimensional crystals can be attributed to the lattice symmetry's mismatching with Mo−d orbitals.

To explicitly verify the nontrivial topological nature of g-MoS$_2$, we have performed the calculations of edge states by cutting 2D monolayer into nanoribbon with the armchair edges. To avoid the interaction between two edges, the width of nanoribbon is cut to be more than 7.40 nm. The module and calculated band structure of g-MoS$_2$ ribbons are presented in Fig. 4. It can be clearly demonstrated that the topologically protected conducting edge states connecting the conduction and valence bands exist within bulk band gap, which confirms the nontrivial topological phase in the g-MoS$_2$. For a 2D QSH insulator, another remarkable characteristic is the $Z_2$ topological invariant (ν), with ν = 1 characterizing a topologically nontrivial phase and ν = 0 meaning a topologically trivial phase. According to the method proposed by Fu and Kane,[52,53] we calculate $Z_2$ topological invariant directly from the parities of Bloch wave functions for occupied energy band at time-reversal-invariant-momenta (TRIM). There are four TRIM points for g-MoS$_2$, namely one $\Gamma$ and three M points. The Z$_2$ topological invariant $\nu$ for g-MoS$_2$ is defined by following equation:

$$\delta(k_i) = \prod_{n=1}^{N} \xi_{2n}^{i}, \quad (-1)^{\nu} = \prod_{i=1}^{4} \delta(k_i) = \delta(\Gamma)\delta(M)^3 \quad (3)$$

Where the $\delta(k_i)$ is the product of parity eigenvalues at the TRIM points, $\xi = \pm 1$ present the parity eigenvalues and N denotes the number of the degenerate occupied energy bands. For g-MoS$_2$, the products of the parity eigenvalues at the $\Gamma$ and M points are −1 and +1, respectively. It implies that the QSH effect can be realized with Z$_2$ topological invariant $\nu = 1$ in g-MoS$_2$ monolayer.

In desirable QSH insulator, large nontrivial gap is one of the prerequisites for practical application in spintronics. To overcome the underestimation of band gaps by the PBE method, we recalculated the band structure of g-MoS$_2$ based on HSE06 funtional.[45,46] Both PBE and HSE06 without SOC show the semi-metal feature with zero band gap. When applying SOC, the degenerated states at the touching point are lifted out (Fig. S5), opening up an energy gap of 40 meV (PBE) and 109 meV (HSE06), much larger than the previous reported TIs materials in 1T′-MoS$_2$ (80 meV),[38] and so-MoS$_2$ (25 meV),[36,37] showing superior performance than graphene, silicone and germanene with nontrivial gaps in the order of meV.[54,55] The relatively large nontrivial band gap makes g-MoS$_2$ promising for practical application as a novel 2D topological insulator at room temperature (~ 30 meV).

Encouragingly, potential gas separation and purification applications in g-MoS$_2$ can be expected. It has been proved that MoS$_2$ nanosheet has a high affinity to selected gas species including H$_2$, NO$_2$ and CO$_2$ due to high adsorption energy of these gas molecules onto the basal surface of MoS$_2$.[56,57] In g-MoS$_2$, the special organization of six- and four-rings renders it is a 2D network with periodically distributed pores. The electron density isosurfaces of porous g-MoS$_2$ monolayer are calculated and shown in Fig. 3c, presenting an average pore diameter of 5.3 Å with a nummular shape via the method described by Song et al.[51] Such a diameter indicates that the porous MoS$_2$ can offer similar pore size distribution as silicalite, and may allow the separation of molecules ($CO_2$, $CH_4$, and $O_2$) based on the differences in diffusion speeds, like other porous materials.[58-60] This feature prompts porous nanosheet a promising material for gas separation and purification applications.

In conclusion, a novel $MoS_2$ allotrope, constituted of $MoS_2$ hexagon and square units, is found to be energetically and thermally stable. Almost all of the contributions near Fermi level come from $d_{z^2}$ orbitals, leading to the existence of heavy fermions. Meanwhile, the nontrivial topological characteristics originated from pure $d$ orbitals are found in this monolayer. The topological features confirmed by edge states and non-zero topological invariant ($v = 1$) show g-$MoS_2$ is a novel QSH insulator with a nontrivial gap of 109 meV, which is large enough for room-temperature detection. The special arrangement of six- and four-rings within the plane endows g-$MoS_2$ the well-defined pore structure, which suggests the potential application in gas separation and storage. Several features, including high stability, quadratic band touching, quantum spin Hall (QSH) effect and special pore structure, open up the possibility for engineering on 2D transition-metal-based QSH insulators, and enable g-$MoS_2$ being promising materials for next-generation high-performance spintronic devices.

## ASSOCIATED CONTENT

**Supporting Information**

Geometry for graphene, *T*-graphene, graphenylene, h-$MoS_2$, so-$MoS_2$ and g-$MoS_2$, *ab-initio* molecular dynamics simulations, shapes of the three atomic $d$ orbitals, band structures with and without SOC, and the Forcite analysis of bond distribution.

## AUTHOR INFORMATION

**Corresponding Authors**

*E-mail: liujiang86@gmail.com.

*E-mail: liming_wu@fjirsm.ac.cn.

**Notes**

The authors declare no competing financial interest.

**Acknowledgment**

This research was supported by the National Natural Science Foundation of China under Projects (21171168, 21225104, 21233009, 21301175, 21571020 and 91422303), the Natural Science Foundation of Fujian Province (2015J01071) and financial support from Bremen University and Max Planck Institute for Chemical Physics of Solids. The supports of the Supercomputer Center of Fujian Institute of Research on the Structure of Matter (FJIRSM) and Supercomputer Center of Northern Germany (HLRN Grant No. hbp00027) are acknowledged.

**REFERENCES**

(1) Novoselov, K. S.; Geim, A. K.; Morozov, S. V.; Jiang, D.; Zhang, Y.; Dubonos, S. V.; Grigorieva, I. V.; Firsov, A. A. Electric field effect in atomically thin carbon films. *Science* **2004**, *306*, 666–669.

(2) Novoselov, K.; Geim, A. K.; Morozov, S. V.; Jiang, D.; Katsnelson, M. I.; Grigorieva, I. V.; Dubonos S. V.; Firsov, A. A. Two-dimensional gas of massless Dirac fermions in graphene. *Nature* **2005**, *438*, 197–200.

(3) Geim, A. K.; Novoselov, K. S. The rise of graphene. *Nat. Mater.* **2007**, *6*, 183–191.

(4) Castro Neto, A. H. C.; Guinea, F.; Peres, N. M. R.; Novoselov K. S.; Geim, A. K. The electronic properties of graphene. *Rev. Mod. Phys.* **2009**, *81*, 109–162.

(5) Vogt, P.; De Padova, P.; Quaresima, C.; Avila, J.; Frantzeskakis, E.; Asensio, M. C.; Resta, A.; Ealet B.; Le Lay, G. Silicene: compelling experimental evidence for graphenelike two-dimensional silicon. *Phys. Rev. Lett.* **2012**, *108*, 155501.

(6) Fleurence, A.; Friedlein, R.; Ozaki, T.; Kawai, H.; Wang Y.; Yamada-Takamura, Y. Experimental

Evidence for Epitaxial Silicene on Diboride Thin Films. *Phys. Rev. Lett.* **2012**, *108*, 245501.

(7) Li, L. F.; Lu, S. Z.; Pan, J. B.; Qin, Z. H.; Wang, Y. Q.; Wang, Y. L.; Cao, G. Y.; Du S. X.; Gao, H. J. Buckled germanene formation on Pt (111). *Adv. Mater.* **2014**, *26*, 4820–4824.

(8) Zhu, F. F.; Chen, W. J.; Xu, Y.; Gao, C. L.; Guan, D. D.; Liu, C. H.; Qian, D.; Zhang, S. C.; Jia, J. f. Epitaxial growth of two-dimensional stanene. *Nat. Mater.* **2015**, *14*, 1020–1025.

(9) Hasan, M. Z.; Kane, C. L. *Colloquium*: topological insulators. *Rev. Mod. Phys.* **2010**, *82*, 3045–3067.

(10) Chen, Y. L.; Kanou, M.; Liu, Z. K.; Zhang, H. J.; Sobota, J. A.; Leuenberger, D.; Mo, S. K.; Zhou, B.; Yang, S. L.; Kirchmann, P. S.; Lu, D. H.; Moore, R. G.; Hussain, Z.; Shen, Z. X.; Qi, X. L.; Sasagawa, T. Discovery of a single topological Dirac fermion in the strong inversion asymmetric compound BiTeCl. *Nat. Phys.* **2013**, *9*, 704–708.

(11) Zhang, H. J.; Liu, C. X.; Qi, X. L.; Dai, X.; Fang Z.; Zhang, S. C. Topological insulators in $Bi_2Se_3$, $Bi_2Te_3$ and $Sb_2Te_3$ with a single Dirac cone on the surface. *Nat. Phys.* **2009**, *5*, 438–442.

(12) Konig, Wiedmann, M.; S.; Brne, C.; Roth, A.; Buhmann, H.; Molenkamp, L. W.; Qi, X. L.; Zhang, S. C. Quantum spin Hall insulator state in HgTe quantum wells. *Science* **2007**, *318*, 766–770.

(13) Knez, I.; Du, R. R.; Sullivan, G. Evidence for helical edge modes in inverted InAs/GaSb quantum wells. *Phys. Rev. Lett.* **2011**, *107*, 136603.

(14) Weng, H. M.; Dai, X.; Fang, Z. Transition-metal pentatelluride $ZrTe_5$ and $HfTe_5$: A paradigm for large-gap quantum spin Hall insulators. *Phys. Rev. X* **2014**, *4*, 011002.

(15) Zhou, L. J.; Kou, L. Z.; Sun, Y.; Felser, C.; F. Hu, M.; Shan, G. C.; Smith, S. C.; Yan, B. H.; Frauenheim, T. New Family of Quantum Spin Hall Insulators in Transition-Metal Halide with Large Nontrivial Band Gaps. *Nano Lett.* **2015**, DOI: 10.1021/acs.nanolett.5b02617.

(16) Song, Z. G.; Liu, C. C.; Yang, J. B.; Han, J. Z.; Ye, M.; Fu, B. T.; Yang, Y. C.; Niu, Q.; Lu, J.; Yao, Y. G. Prediction of Large-Gap Two-Dimensional Topological Insulators Consisting of Bilayers of Group III Elements with Bi. *Nano Lett.* **2014**, *14*, 2505–2508.

(17) Ma, Y. D.; Dai, Y.; Kou, L.; Frauenheim, T.; Heine, T. Robust Two-Dimensional Topological Insulators in Methyl-Functionalized Bismuth, Antimony, and Lead Bilayer Films. *Nano Lett.* **2015**, *15*, 1083–1089.

(18) Ma, Y. D.; Dai, Y.; Wei, W.; Huang, B. B.; Whangbo, M. Strain-induced quantum spin Hall effect in methyl-substituted germanane GeCH$_3$. *Sci. Rep.* **2014**, *4*, 7297.

(19) Ma, Y. D.; Kou, L.; Du, A.; Heine, T. Group 14 element-based non-centrosymmetric quantum spin Hall insulators with large bulk gap. *Nano. Res.* **2014**, *8*, 3412–3420.

(20) Si, C.; Liu, J.; Xu, Y.; Wu, J.; Gu, B.; Duan, W. Functionalized germanene as a prototype of large-gap two-dimensional topological insulators. *Phys. Rev. B* **2014**, *89*, 115429.

(21) Xu, Y.; Yan, B. H.; Zhang, H. J.; Wang, J.; Xu, G.; Tang, P. Z.; Duan, W. H.; Zhang, S. C. Large-gap quantum spin Hall insulators in tin films. *Phys. Rev. Lett.* **2013**, *111*, 136804.

(22) Luo, W.; Xiang, H. J. Room Temperature Quantum Spin Hall Insulators with a Buckled Square Lattice. *Nano Lett.* **2015**, *15*, 3230–3235.

(23) Zhou, J. J.; Feng, W.; Liu, C. C.; Guan, S.; Yao, Y. G. Large-Gap Quantum Spin Hall Insulator in Single Layer Bismuth Monobromide Bi$_4$Br$_4$. *Nano Lett.* **2014**, *14*, 4767–4771.

(24) Song, Z.; Liu, C. C.; Yang, J.; Han, J.; Ye, M.; Fu, B.; Yang, Y.; Niu, Q.; Lu, J.; Yao, Y. Quantum spin Hall insulators and quantum valley Hall insulators of BiX/SbX (X=H, F, Cl and Br) monolayers with a record bulk band gap. *NPG Asia Mater.* **2014**, *6*, e147.

(25) Li, L. Y.; Zhang, X. M.; Chen, X.; Zhao, M. W. Giant Topological Nontrivial Band Gaps in Chloridized Gallium Bismuthide. *Nano Lett.* **2015**, *15*, 1296–1301.

(26) Zhang, H.; Lu, S. B.; Zheng, J.; Du, J.; Wen, S. C.; Tang, D. Y.; Loh, K. P. Molybdenum disulfide (MoS$_2$) as a broadband saturable absorber for ultra-fast photonics. *Opt. Express* **2014**, *22*, 7249–7260.

(27) Zou, X. X.; Zhang, Y. Noble metal-free hydrogen evolution catalysts for water splitting. *Chem. Soc. Rev.* **2015**, *44*, 5148–5180.

(28) Winer, W. O. *Wear*, **1967**, *10*, 422–452.

(29) Mouri, S.; Miyauchi Y. H.; Matsuda, K. Tunable photoluminescence of monolayer $MoS_2$ via chemical doping. *Nano Lett.* **2013**, *13*, 5944–5948.

(30) Mak, K. F.; Lee, C.; Hone, J.; Shan J.; Heinz, T. F. Atomically thin $MoS_2$: a new direct-gap semiconductor. *Phys. Rev. Lett.* **2010**, *105*, 136805.

(31) Conley, H. J.; Wang, B.; Ziegler, J. I.; Haglund, R. F.; Jr.; Pantelides, S. T.; Bolotin, K. I. Bandgap engineering of strained monolayer and bilayer $MoS_2$. *Nano Lett.* **2013**, *13*, 3626–3630.

(32) Kim, S.; Konar, A.; Hwang, W. S.; Lee, J. H.; Lee, J.; Yang, J.; Jung, C.; Kim, H.; Yoo, J. B.; Choi, J. Y.; Jin, Y. W.; Lee, S. Y.; Jena, D.; Choi, W.; Kim, K.; High-mobility and low-power thin-film transistors based on multilayer $MoS_2$ crystals. *Nat. Commun.* **2012**, *3*, 1011.

(33) Radisavljevic, B.; Radenovic, A.; Brivio, J.; Giacometti, V.; Kis, A. Single–layer $MoS_2$ transistors. *Nat. Nanotechnol.* **2011**, *6*, 147–150.

(34) Das, S.; Chen, H. Y.; Penumatcha, A. V.; Appenzeller, J. High performance multilayer $MoS_2$ transistors with scandium contacts. *Nano Lett.* **2012**, *13*, 100–105.

(35) Li, W. F.; Guo, M.; Zhang, G.; Zhang, Y. W. Gapless $MoS_2$ allotrope possessing both massless Dirac and heavy fermions. *Phys. Rev. B* **2014**, *89*, 205402.

(36) Ma, Y. D.; Kou, L. Z; Li, X.; Dai, Y.; Smith, S. C.; Heine, T. Quantum Spin Hall Effect and Topological Phase Transition in Two-Dimensional Square Transition Metal Dichalcogenides. *Phys. Rev. B* **2015**, *92*, 085427.

(37) Sun, Y.; Felser, C.; Yan, B. H. Graphene-like Dirac states and quantum spin Hall insulators in square-octagonal $MX_2$ (*M*= Mo, W; *X*= S, Se, Te) isomers. *Phys. Rev. B* **2015**, *92*, 165421.

(38) Qian, X. F.; Liu, J. W.; Fu L.; and Li, J. Quantum spin Hall effect in two-dimensional transition metal dichalcogenides. *Science* **2014**, *346*, 1344–1347.

(39) Van der Zande, A. M.; Huang, P. Y.; Chenet, D. A.; Berkelbach, T. C.; You, Y. M.; Lee, G. H.; Heinz,

T. F.; Reichman, D. R.; Muller, D. A.; Hone, J. C. Grains and grain boundaries in highly crystalline monolayer molybdenum disulphide. *Nat. Mater.* **2013**, *12*, 554–561.

(40) Najmaei, S.; Liu, Z.; Zhou, W.; Zou, X. L.; Shi, G.; Lei, S. D.; Yakobson, B. I.; Idrobo, J. C.; Ajayan, P. M.; Lou, J. Vapour phase growth and grain boundary structure of molybdenum disulphide atomic layers. *Nat. Mater.* **2013**, *12*, 754–759.

(41) Zhang, Y.; Zhang, Y. F.; Ji, Q. Q.; Ju, J.; Yuan, H. T.; Shi, J. P.; Gao, T.; Ma, D. L.; Liu, M. X.; Chen, Y. B.; Song, X. J.; Hwang, H. Y.; Cui, Y.; Liu, Z. F. Controlled growth of high-quality monolayer $WS_2$ layers on sapphire and imaging its grain boundary. *ACS nano* **2013**, *7*, 8963–8971.

(42) Kresse, G.; Furthmüller, J. Efficient iterative schemes for ab initio total-energy calculations using a plane-wave basis set. *Phys. Rev. B: Condens. Matter Mater. Phys.* **1996**, *54*, 11169–11186.

(43) Kresse G.; Joubert, D. From ultrasoft pseudopotentials to the projector augmented-wave method. *Phys. Rev. B: Condens. Matter Mater. Phys.* **1999**, *59*, 1758–1775.

(44) Blöchl, P. E. Projector augmented-wave method. *Phys. Rev. B: Condens. Matter Mater. Phys.* **1994**, *50*, 17953–17979.

(45) Heyd, J.; Scuseria, G. E. Ernzerhof, M. Hybrid functionals based on a screened Coulomb potential. *J. Chem. Phys* **2003**, *118*, 8207–8215.

(46) Heyd, J.; Scuseria, G. E.; Ernzerhof, M. Erratum:"Hybrid functionals based on a screened Coulomb potential"[J. Chem. Phys. 118, 8207 (2003)]. *J. Chem. Phys* **2006**, *124*, 219906.

(47) Baroni, S.; Gironcoli, S. D.; Corso, A. D.; Giannozzi, P. Phonons and related crystal properties from density-functional perturbation theory. *Rev. Mod. Phys.* **2001**, *73*, 515–562.

(48) Togo, A.; Oba, F.; Tanaka, I. First-principles calculations of the ferroelastic transition between rutile-type and $CaCl_2$-type $SiO_2$ at high pressures. *Phys. Rev. B* **2008**, *78*, 134106.

(49) Nosé, S. A unified formulation of the constant temperature molecular dynamics methods. *J. Chem. Phys* **1984**, *81*, 511–519.

(50) Momma, K.; Izumi, F. *VESTA 3* for three-dimensional visualization of crystal, volumetric and morphology data. *J. Appl. Crystallogr.* **2011**, *44*, 1272–1276.

(51) Song, Q.; Wang, B.; Deng, K.; Feng, X. L.; Wagner, M.; Gale, J. D.; Müllen, K.; Zhi, L. J. Graphenylene, a unique two-dimensional carbon network with nondelocalized cyclohexatriene units. *J. Mater. Chem.* **2013**, *1*, 38–41.

(52) Fu, L.; Kane, C. L. Topological insulators with inversion symmetry. *Phys. Rev. B* **2007**, *76*, 045302.

(53) Fu, L.; Kane, C. L.; Mele, E. J. Topological insulators in three dimensions. *Phys. Rev. Lett.* **2007**, *98*, 106803.

(54) Yao, Y. G.; Ye, F.; Qi, X. L.; Zhang, S. C.; Fang, Z. Spin-orbit gap of graphene: First-principles calculations. *Phys. Rev. B Phys.* **2007**, *75*, 041401.

(55) Liu, C. C.; Feng, W. X.; Yao, Y. G. Quantum spin Hall effect in silicene and two-dimensional germanium. *Phys. Rev. Lett.* **2011**, *107*, 076802.

(56) Berean, K. J.; Ou, J. Z.; Daeneke, T.; Carey, B. J.; Nguyen, E. P.; Wang, Y. C.; Russo, S. P.; Kaner, R. B.; Zadeh, K. K. 2D $MoS_2$ PDMS Nanocomposites for $NO_2$ Separation. *Small* **2015**, *38*, 5035–5040.

(57) Wang, D.; Wang, Z. G.; Wang, L.; Hu L.; Jin, J. Ultrathin membranes of single-layered $MoS_2$ nanosheets for high-permeance hydrogen separation. *Nanoscale* **2015**, *7*, 17649–17652.

(58) Bernardo, P.; Drioli, E.; Golemme, G. Membrane gas separation: a review/state of the art. *Ind. Eng. Chem. Res.* **2009**, *48*, 4638–4663.

(59) Zornoza, B.; Esekhile, O.; Koros, W. J.; Téllez, C.; Coronas, J. Hollow silicalite-1 sphere-polymer mixed matrix membranes for gas separation. *Sep. Purif. Technol.* **2011**, *77*, 137–145.

(60) Mahmood, J.; Lee, E. K.; Jung, M.; Shin, D.; Jeon, I. Y.; Jung, S. M.; Choi, H. J.; Seo, J. M.; Bae, S. Y.; Sohn, S. D.; Park, N.; Oh, J. H.; Shin, H. J.; Baek, J. B. Nitrogenated holey two-dimensional structures. *Nat. Commun.* **2015**, *6*, 6486.

# Figure Caption

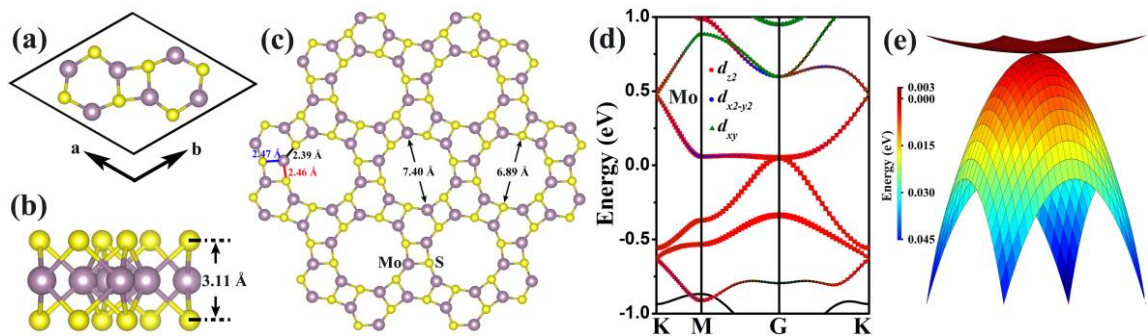

Figure 1. Equilibrium structure of g-MoS$_2$ from (a) top view and (b) side view. (c) Hexagonal model representation. Primitive unit cells are emphasized by the solid black line. (d) Calculated orbital-resolved band structures of g-MoS$_2$ based on DFT-PBE. (e) 3D band structures formed by the valence and conduction band. The Fermi level has been set at 0 eV. The bands near Fermi energy are mainly dominated by Mo−$d_{z2}$, Mo−$d_{x2-y2}$ and Mo−$d_{xy}$.

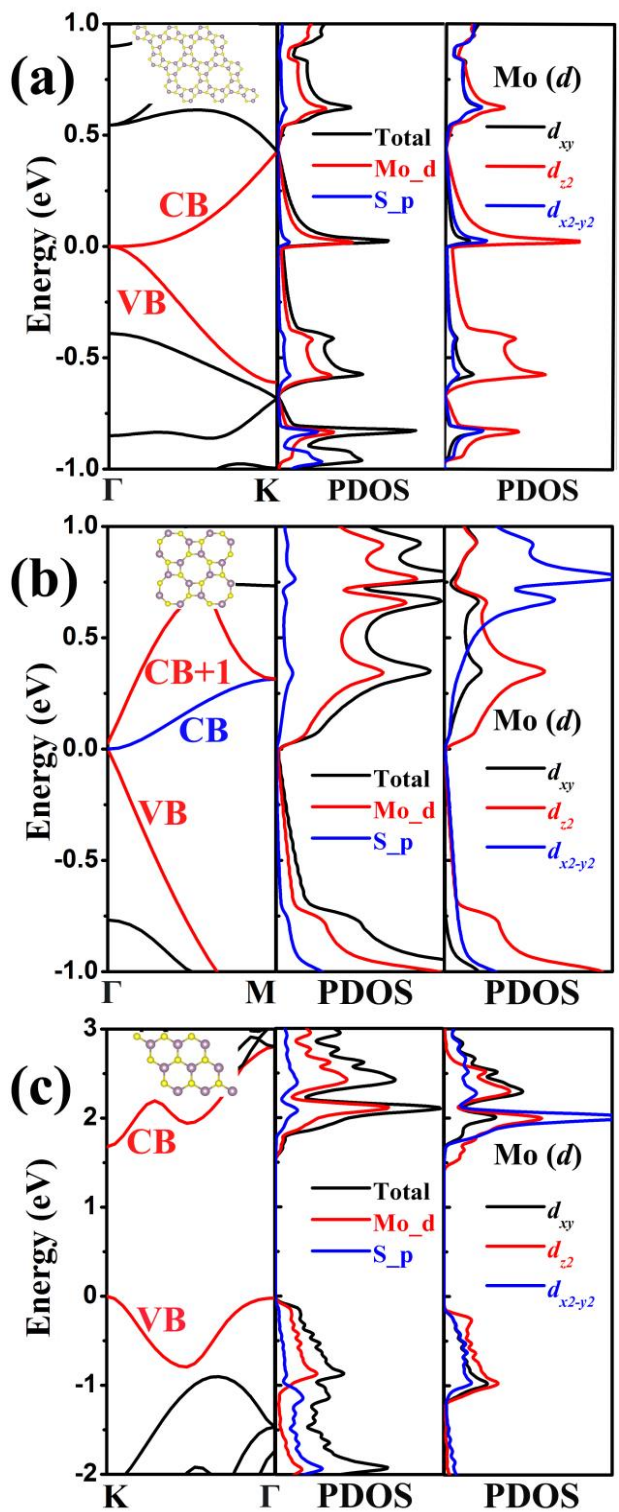

Figure 2. The selected (Γ − k, Γ − M and K − Γ) band structures and density of states for (a) g-MoS$_2$, (b) so-MoS$_2$ and (c) h-MoS$_2$ monolayer.

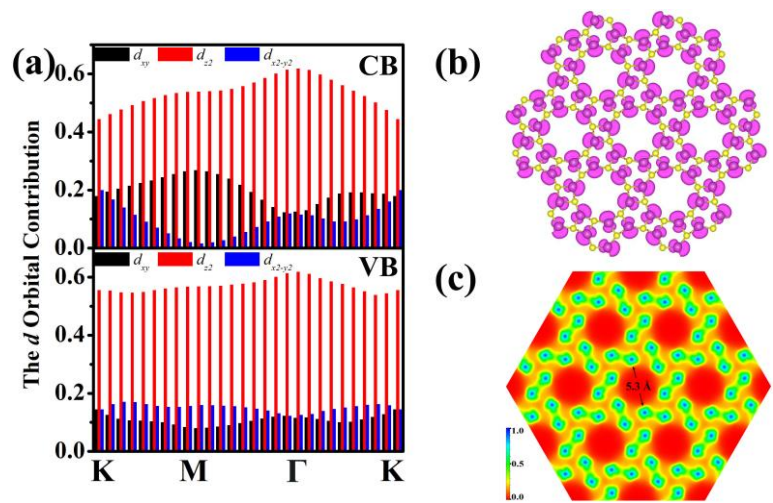

Figure 3. (a) The contributions of three main *d* orbitals to VB (down) and CB (top). (b) VBM charge density contours at Γ point of g-MoS$_2$. (c) Electron density isosurfaces for g-MoS$_2$. Isosurface value: 0.002 e/Å$^3$.

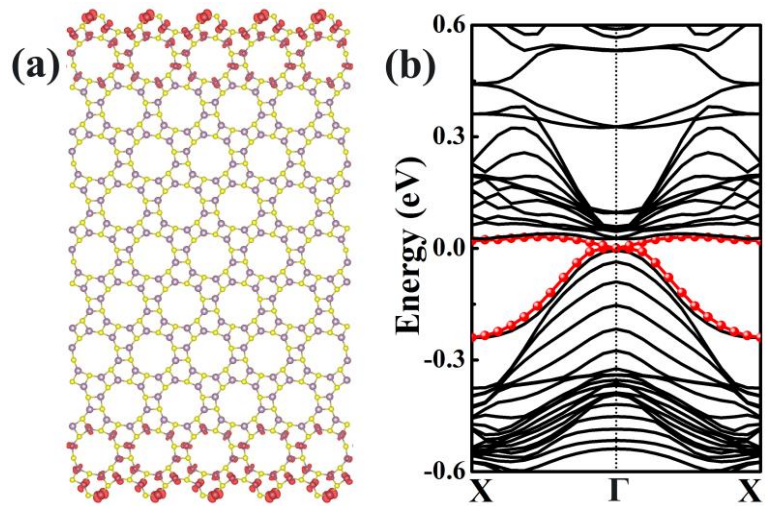

Figure 4. (a) Top view of g-MoS$_2$ nanoribbon with armchair edges. (b) The calculated topological edge states for g-MoS$_2$ with SOC. The Dirac helical states are denoted by the red solid lines and red spheres, which exist at the edges of the ribbon structure. The partial charge density for helical states bands (in $10^{-3}$ e/Å). Zero of energy is set at the Fermi level.